\documentstyle[a4,12pt,psfig]{article}
\begin{document}

\newcommand{\al}{\alpha}
\newcommand{\bt}{\beta}
\newcommand{\1}{\vspace{1cm}\noindent}
\newcommand{\4}{\vspace{4cm}\noindent}
\newcommand{\be}{\begin{equation}}
\newcommand{\ee}{\end{equation}}
\newcommand{\ba}{\begin{eqnarray}}
\newcommand{\ea}{\end{eqnarray}}
\newcommand{\de}{\delta}
\newcommand{\dd}{\partial}
\newcommand{\ga}{\gamma}
\newcommand{\sg}{\sigma}
\newcommand{\db}{\bar{\partial}}
\newcommand{\fb}{\bar{f}}
\newcommand{\tet}{e_z^{~~a}}
\newcommand{\tec}{e_{\bar{z}}^{~~a}}
\newcommand{\teo}{e_0^{~~a}}
\newcommand{\ta}{\theta}
\newcommand{\ach}{\mbox{arccosh}}
\newcommand{\bz}{\bar{z}}
\newcommand{\ra}{\rightarrow}
\newcommand{\la}{\lambda}
\newcommand{\nn}{\nonumber}
\newcommand{\p}{\psi^\al(z)}
\newcommand{\pbar}{\bar{\psi}_\beta(z_o)}
\newcommand{\vs}{\vspace{0.2cm}}
\newcommand{\vsa}{\vspace{0.4cm}}
\newcommand{\vsb}{\vspace{0.8cm}}
\newcommand{\ha}{\frac{1}{2}}
\newcommand{\eps}{\varepsilon}
\newcommand{\vfi}{\varphi}
\newcommand{\dr}{e^{~~a}_\mu}
\newcommand{\scon}{\omega^{~~a}_\mu}
\newcommand{\pijl}{\leftrightarrow}
\newcommand{\cd}{{\cal D}}
\newcommand{\cb}{{\cal B}}
\newcommand{\cc}{{\cal C}}
\newcommand{\veps}{\varepsilon}
\newcommand{\vx}{\vec{x}}
\newcommand{\vp}{\vec{p}}
\newcommand{\om}{\omega}
\newcommand{\ck}{{\cal K}^l_{m,(t)}(\xi,\vfi)}
\newcommand{\cdd}{\cd^l_{m,t}(H,\vfi,\xi)}
\newcommand{\vt}{\vartheta}
\newcommand{\Ml}{{\cal M}^\la_m(\eta,\vt)}
\newcommand{\Mr}{\overline{{\cal M}^{\la}_{m}}(\eta,\vt)}
\newcommand{\Mrs}{\overline{{\cal M}^{\la'}_{m'}}(\eta,\vt)}
\newcommand{\Mrr}{\overline{{\cal M}^{\la}_{m}}(\eta',\vt')}

\begin{titlepage}
\begin{flushright}
Preprint: THU-97/10\\
hep-th/9704067 \\
April 1997
\end{flushright}
\vsa
\begin{center}
{\large\bf Winding Solutions for the two Particle System\vs\\
           in 2+1 Gravity\vsa\vsb\\}
           M.Welling\footnote{E-mail: welling@fys.ruu.nl\\
                     Work supported by the European Commission TMR programme
                                                         ERBFMRX-CT96-0045}
     \vsa\vsb\\
   {\it Instituut voor Theoretische Fysica\\
     Rijksuniversiteit Utrecht\\
     Princetonplein 5\\
     P.O.\ Box 80006\\
     3508 TA Utrecht\\
     The Netherlands}\vsb\vsa\\
\end{center}
\begin{abstract}
Using a PASCAL program to follow the evolution of two gravitating particles
in 2+1 dimensions we find solutions in which the particles wind around one
another indefinitely. As their center of mass moves `tachyonic' they form
a Gott-pair. To avoid unphysical boundary conditions we consider a large but
closed universe. After the particles have evolved for some time their
momenta have grown very large. In this limit we quantize the model and
find that both the relevant configuration variable {\em and} its
conjugate momentum become discrete.
\end{abstract}
\end{titlepage}

\section{Introduction}
7 Years after the paper of Deser, Jackiw and 't Hooft on 2+1 dimensional
gravity \cite{DJH}, Gott published an article in which he constructed a space
time containing closed time like curves (CTC) without introducing `exotic
matter' \cite{Gott}. The basic idea was that two cosmic strings in 3+1
dimensional gravity, or particles in 2+1 dimensional gravity, approaching each
other with high velocity and small impact parameter could produce these CTC's.
It was however recognized quickly thereafter that the CTC is not confined to a
small region of space time but also exists at spatial infinity \cite{DJH2}.
This fact implies that Gott's space time violates `physical boundary
conditions' that should be imposed at infinity. Another way of saying it is
that the center of mass (c.o.m.) of the two particles is `tachyonic', although
the strings themselves are particlelike. Tachyonic means that the energy
momentum vector is spacelike. So physical reasonable boundary conditions mean
that the total energy momentum of the universe is timelike. Next, Cutler showed
that one can find a family of Cauchy surfaces, prior to the existence of the
CTC's, from which they evolve \cite{Cutler}. This implies that evolution of
initial data from these Cauchy surfaces must run into a Cauchy Horizon. Another
interesting question that arose was whether it was possible to start with a
space time of timelike particles and accelerate two particles to high
velocities in such a way that they form a Gott-pair. It was found in
\cite{Guth} that there is never enough energy in an open universe to achieve
this. In a closed universe however the Gott-pair can be formed and initially it
{\em seemed} possible to construct CTC's again. But now the `chronology
protection conjecture' \cite{Hawking} was saved by a very different effect. 't
Hooft showed that a closed universe containing Gott-pairs will crunch before
the CTC's can be fully traversed. Once the Gott-pair has come into existence
the cosmic strings start to spiral around one another with ever increasing
velocity
until they reach the `end of the universe' causing it to crunch.

In this paper we reproduce this winding solution (in a slightly different way)
using 't Hooft's Cauchy formulation. Despite the fact that the strings move
subluminally, the c.o.m. moves `tachyonic', i.e. faster than the speed of
light. If we want to avoid the unphysical asymptotic boundary conditions we
must close the universe, following the philosophy of 't Hooft:\\
\\
{\em "A safe way to consider open universes is to view them as limiting cases
of infinitely large closed spaces."}\\
\\
This means that we start with one of Cutler's Cauchy surfaces and evolve the
initial data from it until we approach the Cauchy Horizon. But in the
philosophy explained above this really implies that the universe has crunched
before we arive at this horizon. This big crunch will happen in the very far
future however if we choose our universe big enough.

In the last section we study the Hilbert-space of the two particles (treating
the rest of the universe at `infinity' classically) and find that it is {\em
finite dimensional}.

 \section{Two Particles described by Polygon Variables}

\begin{figure}[t]
\centerline{\psfig{figure=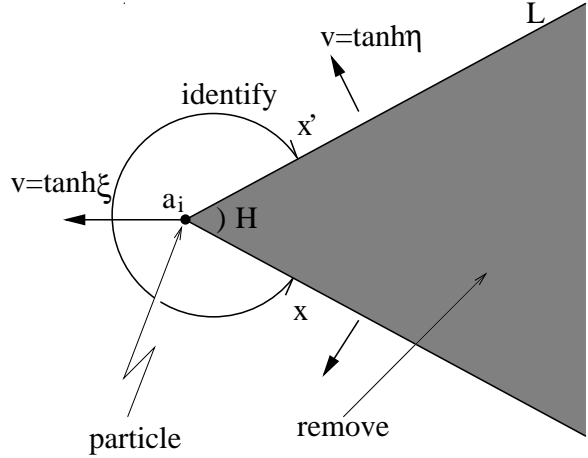,angle=-90,height=6cm}}
\caption{Moving particle with a wedgelike region cut out.}
\label{wedges}
\end{figure}

\begin{figure}[t]
\centerline{\psfig{figure=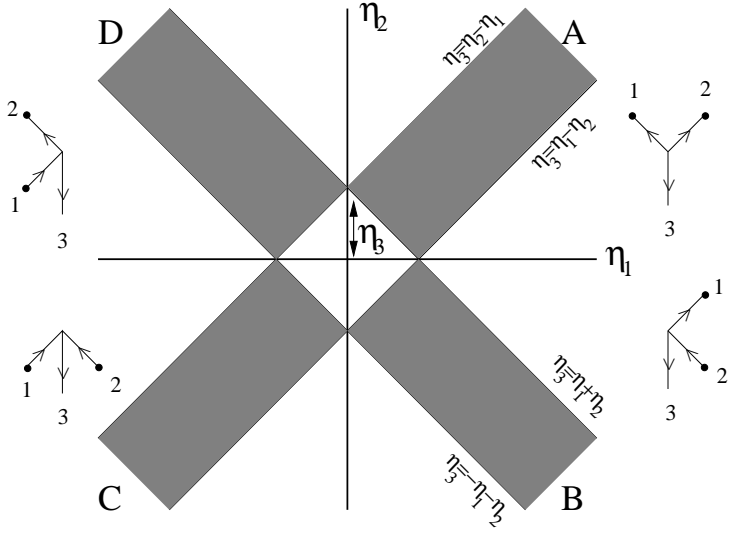,angle=-90,height=7cm}}
\caption{Allowed values for $\eta_1$ and $\eta_2$ at a fixed value for
$\eta_3$.}
\label{momenta}
\end{figure}

A simple way to describe a particle with mass $M$ in 2+1 dimensional gravity
was already discovered by Staruszkiewicz in 1963 \cite{Star}. Let's suppose
this particle is sitting at the location ${\bf a}$. The effect of its
gravitational field is described by cutting out a wedge from space time and
identifying the boundaries. The angle of the excised region is $8\pi GM$. From
now on we will set $8\pi G=1$. The identification is done by a simple rotation:
\be
{\bf x'}={\bf a}+R(M)({\bf x-a})
\ee
A moving particle is described by boosting this solution, using a boostmatrix
$B(\xi)\in$ SO(2,1) (see figure \ref{wedges}). The velocity of the particle is
$v=\tanh\xi$.
Now we have:
\be
{\bf x'}={\bf a}+B(\xi)~R(M)~B^{-1}(\xi)({\bf x-a})
\ee
Notice that if we choose the wedge behind (or in front) of the particle the
identification rule has no time jump. By Lorentz contraction the excised angle
is larger than is the case for a particle at rest. It turns out that this total
angle represents the energy of the particle. It is given by:
\be
\tan\frac{H}{2}=\cosh\xi~\tan\frac{M}{2}
\ee
The fact that the energy grows as we enhance the velocity of the particle makes
sense because we add kinetic energy to the particle. The next step is to
describe two particles at positions ${\bf a_1}$ and ${\bf a_2}$ with arbitrary
velocities ${\bf v_1}$ and ${\bf v_2}$. Now cut out the wedges in front or
behind the particles in such a way that they meet somewhere. The situation is
pictured in figure (\ref{2deeltjes}). The point where the boundaries of the
wedges meet is called a vertex. The idea is now that after this vertex we
continue our cutting in such a way as if there where an effective center of
mass (c.o.m.) particle with a certain speed and spin (which equals the total
angular momentum of the system). Consider the point ${\bf b_1}$ and ${\bf b_2}$
in the figure. They are related by the identification rule:
\be
{\bf b_1}=B_1~R_1~B_1^{-1}[\{B_2~R_2~B_2^{-1}({\bf b_2-a_2})+{\bf a_2}\}
-{\bf a_1}]+{\bf a_1}\label{id1}
\ee
This can be written as:
\be
{\bf b_1}=B_T~R_T~B_T^{-1}({\bf b_2-d})+{\bf d}+B_T~{\bf\ell}
\label{id2}
\ee
where ${\bf d}$ is the location of the c.o.m. particle, $B_T$ is the total
boost determined by the velocity of the c.o.m. particle, $R_T$ is a rotation
over an angle representing the energy of the system in the c.o.m. frame and
${\bf\ell}=(l,0,0)$ where $l$ is the total angular momentum of the system in
the c.o.m. frame (see \cite{Welling3}).
By comparing (\ref{id1}) and (\ref{id2}) we can express these quantities in the
parameters of the individual particles. If we demand that there are no time
jumps in (\ref{id2}), i.e. $t_1=t_2$, we can calculate the location of the
boundary
of $L_3$. The angle of this wedge represents the total energy of this system
$H_T$. The total angular momentum is represented as a spatial shift over the
boundary. We mention that we can only avoid time jumps if we are not in the
c.o.m. frame. We will introduce the polygon variables $\eta_i$ and $L_i$ in the
following. $L_i$ is given by the length of the wedge behind (or in front) of
particle $i$. $L_3$ is taken to be the length of the boundary of the total
wedge. $\eta_i$ is the rapidity ($v_L=\tanh\eta$) of the boundary which moves
perpendicular to itself (see figure (\ref{wedges})). We can express the total
energy in terms of $\eta_i$ only. It is given by:
\begin{figure}[t]
\centerline{\psfig{figure=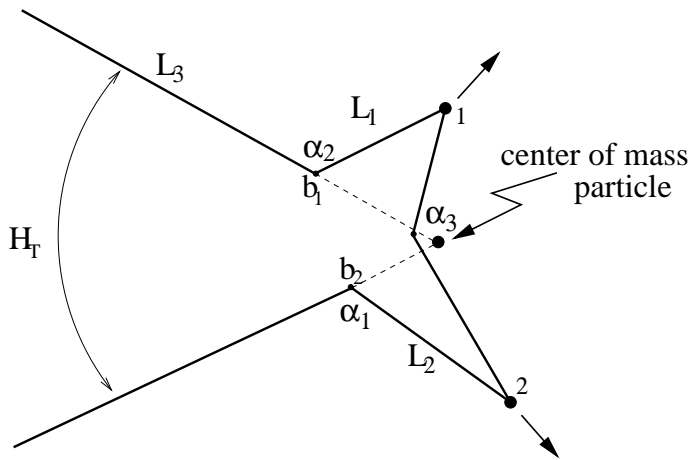,angle=-90,height=6cm}}
\caption{Two particles which cut out a piece of space time.}
\label{2deeltjes}
\end{figure}
\be
H_T=H_1+H_2+2\pi-\al_1-\al_2-\al_3\label{Ham}
\ee
where
\be
H_i=2\arccos(\frac{\cos\frac{M_i}{2}}{\cosh\eta_i})
\ee
and
\ba
\cos\al_1&=&\frac{\ga_1-\ga_2\ga_3}{\sg_2\sg_3}\label{al1}\label{a1}\\
\cos\al_2&=&\frac{\ga_2-\ga_1\ga_3}{\sg_1\sg_3}\label{al2}\label{a2}\\
\cos\al_3&=&\frac{\ga_3-\ga_1\ga_2}{\sg_1\sg_2}\label{al3}\label{a3}
\ea
where we have defined:
\be
\ga_i=\cosh(2\eta_i)~~~~~~~~\sg_i=\sinh(2\eta_i)
\ee
The relation among the angles $\al_i$ and momenta $\eta_j$ can be calculated
using the fact that there is no mass at the vertex, so space time should be
flat there. We did however introduce a two dimensional curvature at the vertex
as the angles $\al_i$ do in general not add up to $2\pi$. This then must be
compensated by extrinsic curvature to produce a flat three dimensional space
time. So if we move a Lorentz vector around the vertex it should not be Lorentz
transformed after returning to its original position. This is expressed as
follows:
\be
I=R(\al_2)~B(2\eta_3)~R(\al_1)~B(2\eta_2)~R(\al_3)~B(2\eta_1)
\label{rondje}
\ee
{}From this we can deduce vertex relations \cite{Hooft1} which enable us to
calculate for instance the angles $\al_i$ from the momenta $\eta_i$. If we use
$H_T(\eta_i)$ we can verify that $L_i$ and $2\eta_i$ are canonically conjugate
variables, i.e.:
\ba
\frac{d}{d t}L_i&=&\{H_T,L_i\}=\ha\frac{\dd H_T}{\dd \eta_i}\\
\frac{d}{d t}\eta_i&=&\{H_T,\eta_i\}=-\ha\frac{\dd H_T}{\dd L_i}=0
\ea
It will furthermore be important that the $\eta_i$ fulfill a triangle
inequality:
\be
|\eta_i|+|\eta_j|\geq |\eta_k|~~~\mbox{and cyclic permutations}
\ee
Let us choose a global Lorentz frame, i.e. we fix the velocity of the c.o.m.
frame.
This amounts to fixing the value of $\eta_3$. What are now the possible values
of $\eta_1$ and $\eta_2$ if we take the triangle inequalities into account
properly. This is best seen in figure (\ref{momenta}).
The coloured region contains the allowed values for $\eta_i$. A negative sign
for $\eta$ means that the edges are moving inward instead of outward. This is
indicated with an arrow: if the arrow is pointing away from the vertex the sign
of $\eta$ is positive and vice versa. We may use as a rule of thumb that the
arrows considered as vectors must add up to zero.
If all the signs of the $\eta_i$ are the same (i.e. the same as the fixed sign
of $\eta_3$) then all angles $\al_i$ are smaller then $\pi$ and we are in
region A. If one of the signs differs from its neighbours, the opposite angle
$\al$ is larger than $\pi$ as is the case in region B,C and D. We chose a fixed
orientation for particle 1 and 2 but the opposite orientation is of course also
possible. If the system evolves, two kind of transitions can occur if one of
the lengths $L_i$ shrinks to zero. The transitions between C and B or D are
easy and orientation preserving. We will call them type-1 transitions in the
following. Let us take for instance the transition from C to B. In this case
only the sign of $\eta$ changes while absolute value remains unchanged. The
values and signs of $\eta_2$ and $\eta_3$ remain fixed. This implies that
$\al_2\ra\al_2+\pi$ and $\al_3\ra\al_3-\pi$ according to the 'sign rule' above.
A more interesting transition occurs between region A and B (or D). We will
call these transitions type-2 in the following. Let us consider a transition
from A to B. Because this transition changes orientation we have to interchange
particle 1 and 2 in region B. As particle 1 hits the edge it appears on the
other side, but its speed is boosted according to:
\be
\ga_1'=\ga_2~\ga_3+\sg_2~\sg_3~\cos(\al_2+\al_3-H_1)
\label{trans}
\ee
where $\al_2$, $\al_3$ and $H_1$ are given in terms of $\eta_i$.
The values and signs of $\eta_2$ and $\eta_3$ remain unchanged. It seems
strange that particle 1 is boosted while the momentum of particle 2 stays the
same. Where does the energy come from? The answer is that there is
gravitational energy stored in the three-vertex which is transformed into
kinetic
energy for particle 1. (\ref{trans}) is one of the transition rules that can be
derived from (\ref{rondje})
\cite{Hooft1}. After the new rapidities have been calculated we can use
(\ref{al1},\ref{al2},\ref{al3}) to calculate the new angles.

Summarizing we can say that the system evolves for a while linearly (the $L_i$
change linear in time) after which a transition takes place. This process may
repeat itself for some time. For low energy scattering we expect only a few
transitions to occur.
For high energy processes however we will find that the particles can wind
around each other indefinitely.

\section{The winding Solution}

In order to follow the evolution of this two particle system we wrote a program
in PASCAL. Typically one starts in region C of figure (\ref{momenta}) where
both particles move towards the vertex. Depending on whether particle 1 or
particle 2 hits the vertex first we make a type-1 transition into region B or D
respectively. After that we move into sector A by a type-2 transition. For low
energies the particles, once they are in region A, move apart linearly in time,
i.e. they are scattered. Because of the deficit angles these particles are
deflected by some scattering angle. If we add more energy to the system we also
find solutions that wind around one another for some time after which they
break free and move apart. If we add so much energy to the system that the
total energy exceeds $\pi$ (but is less than $2\pi$) we find that there are
many solutions that wind around each other
indefinitely. In other words, the particles are trapped in their own
gravitational field.

Let us briefly describe how the computer program works. First it determines in
what sector the particles are. It then calculates the values of $\dot{L}_1$ and
$\dot{L}_2$. It is important that $\dot{L}_i$ has two contributions, one from
the fact that the particles move through Minkowski space and one from the fact
that the vertex moves. For $\dot{L}_1$ we find:
\ba
\dot{L}_1 &=& \ha\frac{\dd H_T}{\dd \eta_1}=\ha\frac{\dd H_1}{\dd
\eta_1}-\ha\frac{\dd}{\dd\eta_1}(\al_1+\al_2+\al_3) \label{V1} \\
          &=& \frac{\cos(\frac{M_1}{2})~\tanh\eta_1}{\sqrt{\cosh^2\eta_1-
\cos^2\frac{M_1}{2}}}+{\bf sign}(\sin\al_2)~\sqrt{\frac{\ga_1-1}{\ga_1+1}}~
\frac{(1+\ga_1-\ga_2-\ga_3)}{\sqrt{1-\ga_1^2-\ga_2^2-\ga_3^2+2\ga_1\ga_2\ga_3}}
\nn
\ea
The formula for $\dot{L}_2$ can be obtained from (\ref{V1}) by interchanging
all indices 1 and 2.
The first term is clearly due to the velocity of the particle itself. Notice
that for large values of $\eta_1$ this velocity is suppressed like
$\dot{L}_1\sim e^{-\eta_1}$. The second term is the contribution from the
vertex. For large $\eta_{1,2}$ this term survives and its value depends on the
difference $(\eta_1-\eta_2)$ (see (\ref{V1'},\ref{V2'})). We find that for
small values of $\eta_i$ only the first term is important and for large values
the second term dominates.

After the program has calculated the velocities $\dot{L}_i$ it determines what
kind of transition will occur. In the case of a type-1 transition it only
changes the sign of the $\eta$ that is involved in the transition. Furthermore
it changes two angles by $\pi$ as was explained in section 1. In case of a
type-2 transition it uses equation (\ref{trans}) to calculate the new value for
$\eta$ (or $\ga$). Of course all angles $\al_i$ change in such a way that the
total energy is conserved. This change is calculated using
(\ref{a1},\ref{a2},\ref{a3}). As the kinetic energy of the particle is
increased
some energy from the vertex is transferred to the particle that is boosted.
After this it reevaluates the velocities and determines whether they will move
to infinity or wind yet another time.

\begin{figure}[t]
\centerline{\psfig{figure=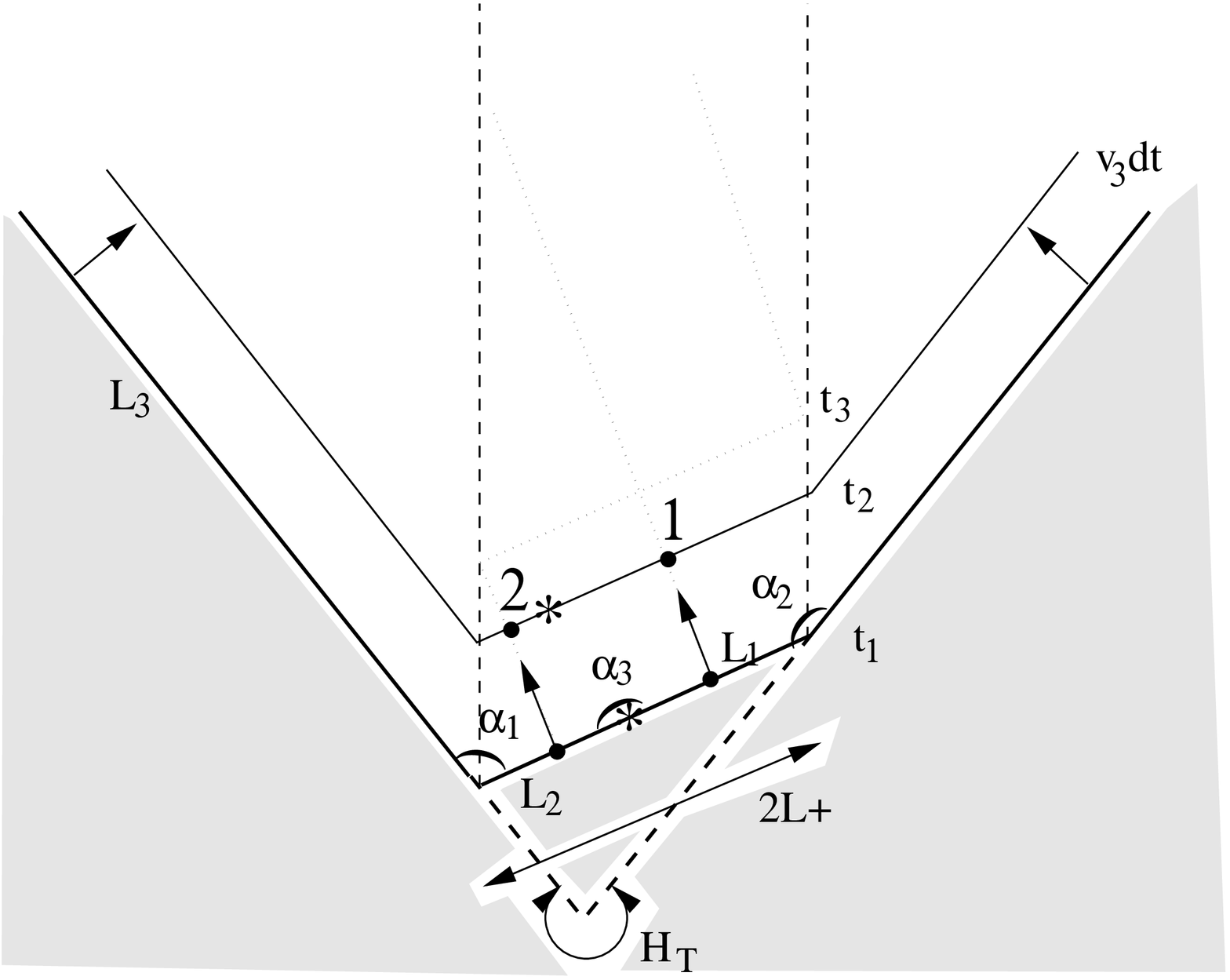,angle=-90,height=8cm}}
\caption{Two particle universe in the limit $p_+\ra\infty$.
The shaded area is removed from space.}
\label{universe}
\end{figure}

Now let's assume we are in sector A, $\dot{L}_1<0$ and $\dot{L}_2>0$. The next
transition will involve particle 1. After the transition we have two
possibilities; either $\dot{L}_1>0$ and $\dot{L}_2>0$ which implies that the
particles break free, or $\dot{L}_1>0$ and $\dot{L}_2<0$ which implies that
they will wind another time. As it turns out, once we are in region A and we
choose $\eta_{1,2}$ large enough and not equal, we have solutions that wind
indefinitely. We will now investigate this behaviour closer. First we introduce
new phase space variables in the following way:
\ba
p_\pm &=&\eta_1\pm\eta_2\\
L_\pm &=&L_1\pm L_2
\ea
If we substitute these variables in the expression for the total Hamiltonian
(\ref{Ham}) and take the limit for large $\eta_{1,2}$, i.e. $p_+\gg 1$, we
find:
\be
H_T=3\pi-\arccos(\frac{e^{2p_-}-\ga_3}{\sg_3})-
\arccos(\frac{e^{-2p_-}-\ga_3}{\sg_3})+{\cal O}(e^{-\ha p_-})
\ee
or
\be
\cos\frac{H_T}{2}=-\frac{\cosh p_-}{\cosh\eta_3}+{\cal O}(e^{-\ha p_-})
\ee
First off all we notice that $H_T$ is inevitably in the range $H_T\in
[\pi,2\pi]$. Secondly, the first term that contains $p_+$ is exponentially
suppressed! It is also important to keep in mind that $p_-$ is restricted
by the triangle inequalities to the range:
\be
p_-\in [-\eta_3,\eta_3]
\ee
as can be seen in figure (\ref{momenta}) by rotating it over 45 degrees. We
also see that in region A we have:
\be
p_+\in [\eta_3,\infty)
\ee
The fact that $p_-$ is restricted to a `Brillouin zone' will cause quantization
of $L_-$ in the quantum theory. Let us now return to the transitions. In the
case of type-2  we find the following result in the limit $p_+\gg 1$:
\be
p_-'=-p_-+{\cal O}(e^{-\ha p_+})\label{trans-}
\ee
So once we are in region A the transition will not kick us out of the allowed
range. We can also deduce what $p_+$ does:
\be
p_+'=p_++2|p_-|+{\cal O}(e^{-\ha p_+})
\ee
As the $\eta_i$ grow, the actual speed of the particles increases until it
reaches the speed of light: $v_i\simeq 1$. In the limit for $p_+\gg 1$ we have
for
$\dot{L}_i$:
\ba
\dot{L}_1&=&\frac{\sinh p_-}{\sqrt{\cosh^2\eta_3-\cosh^2 p_-}}+{\cal O}(e^{-\ha
p_+})
\label{V1'}\\
\dot{L}_2&=&\frac{-\sinh p_-}{\sqrt{\cosh^2\eta_3-\cosh^2 p_-}}+{\cal
O}(e^{-\ha p_+})
\label{V2'}\\
\dot{L}_3&=&\frac{-\cosh p_-}{\sqrt{\cosh^2\eta_3-\cosh^2 p_-}}+{\cal
O}(e^{-\ha p_+})
\ea
We notice that:
\be
\dot{L}_+=0+{\cal O}(e^{-\ha p_+})
\ee
This was to be expected as in this limit $H_T$ only depends on $p_-$. All
dependence on $p_+$ is exponentially suppressed as $p_+$ grows. This implies
that we may view $L_+$ as a fixed parameter in this limit. Its value depends on
the initial condtitions. The only configuration variable to survive is $L_-$
and we find:
\be
\dot{L}_- =\frac{2\sinh p_-}{\sqrt{\cosh^2\eta_3-\cosh^2 p_-}}+{\cal O}(e^{-\ha
p_+})\label{V-}
\ee
Next we turn to construct a geometric picture of what is going on in this
limit.
Therefore we calculate all angles $H_i$ and $\al_i$:
\ba
H_{1,2}&=&\pi+{\cal O}(e^{-\ha p_+})\\
\al_3&=&\pi+{\cal O}(e^{-p_+})\\
\al_{1,2}&=&\arccos(\frac{e^{\pm 2p_-}-\ga_3}{\sg_3})+{\cal O}(e^{-2 p_+})
\ea
Putting this information together we construct a picture of (the finite part
of) our universe: see
figure (\ref{universe}).
Firstly we notice that $H_T\in [\pi,2\pi]$. As $H_{1,2}$ and $\al_3$ are equal
to $\pi$ it follows that $L_1$ and $L_2$ form a straight line. The total length
of this line is $2 L_1+2 L_2=2 L_+$. As we have seen this must be a constant in
this limit. The distance between the particles is equal to $L_+$. We also know
that $L_3$ moves perpendicular to itself with a velocity $v_3=\tanh\eta_3$ and
the particles move with the speed of light perpendicular to the above mentioned
line ($\eta_{1,2}\ra\infty,~\xi_i\ra\infty$). Using this information we can
construct the location of the $L_i$ and the particle's positions at a time step
later. The angles $\al_i$ remain constant because they depend only on the
momenta $\eta_i$. At $t=t_2$ we see that the location of the * has moved closer
to particle 2. This implies that $L_-=L_1-L_2$ has grown according to
(\ref{V-}). If we let the system evolve to $t=t_3$, $L_2$ has shrunk to zero
and must make a transition of type-2 (\ref{trans-}). We see from the picture
that the particle simply reappears on the other side of the line. It is as if
the particles were moving on a circle. This will also become important in the
quantum theory as it implies quantization of $p_-$!

It is important to notice that after each transition $p_+\ra p_++2|p_-|$, so
the approximation becomes better and better as the system evolves in time. Once
we have entered the region in phase space for which $p_+\gg 1$, evolution only
drags us in further.

One can easily see from the figure that the c.o.m. moves tachyonic, or faster
than the speed of light. As the individual particles move (almost) with the
speed of light it follows that the vertex-points at $\al_1$ and $\al_2$ move
superluminal along the vertical dashed lines. Because the angles themselves do
not change this is the same speed as the tip of the cone (i.e. the
c.o.m.-tachyon). The fact that this is possible while the particles move
subluminal is caused by the instant jumps during a transition along the
vertical dashed lines. One can verify this by a simple calculation:
\ba
v_{\bf com}&=&\frac{\tanh\eta_3}{\sin\frac{H}{2}}\\
           &=&\frac{\sinh\eta_3}{\sqrt{\cosh^2\eta_3-\cosh^2p_-}}\geq
           1~~~~\mbox{if}~~~~p_-\in [-\eta_3,\eta_3]
\ea
So we must conclude that we formed a Gott-pair! Because we used a Cauchy
formulation we do not have to worry about CTC's. As Cutler showed, a family of
Cauchy surfaces exists before the appearance of the CTC's. But then we do have
to worry about the fact that our Cauchy formulation runs into a Cauchy Horizon
and we cannot integrate past this point\footnote{I thank H.J. Matschull for
pointing this out to me}. At this Cauchy Horizon the first closed lightlike
curve comes into existence. In that case we have chosen an `unphysical'
universe. This happens for instance in an open universe with a Gott-pair.
Carroll, Farhi and Guth have shown that there is simply not enough energy in
the (open) universe to create such a pair from regular initial conditions. So
then one must conclude that tachyons have created the pair and as we don't
observe these tachyons we exclude this possibility. Another way of stating this
is that the identification at infinity is boostlike instead of rotationlike.
This in turn implies that the total energy momentum vector is spacelike. This
boundary condition was characterized as unphysical in \cite{DJH2}. A possible
way to avoid the problem of unphysical boundary conditions is to close our
universe `at infinity' by adding some `spectator particles'. It was shown that
in the literature \cite{Guth} that the Gott-pair can be formed safely in a
closed universe. The fact that no CTC's appear in 't Hooft's description is due
to the fact that before they can be traversed the Gott-pair has reached the
spectator particles at `infinity'. 't Hooft showed that the universe then
necessarily crunches \cite{Hooft2}. So the Cauchy Horizon is screened off by
the big crunch. But we can postpone this dramatic ending of the universe by
simply locating the observer-particles very far away. Before the crunch the
spectator particles have plenty of time to study the two particles. This will
be the philosophy in the next section. The particles that close the universe at
infinity are treated classically. They have clocks that define time in the
universe. They study the Hilbert-space of the two particles that approach them
superluminally. This will however be a difficult job because the light that
reaches them has necessarily undergone lots of transitions themselves (light
that travels in straight lines will be overtaken by the c.o.m.-tachyon). Notice
that a local observer also necessarily makes a lot of transitions and is
boosted to very high velocities. Due to time dilatation he may observe the two
particles very differently. An important lesson is that these states cannot
exist in open universes with physical boundary conditions (which makes them
unimportant for scattering calculations), but only exist in closed universes
(although they may be very large). The final crunch in this universe is however
unavoidable as it must screen off the potential CTC's after the Cauchy Horizon.

\section{Quantum Theory}
In this section we will consider the quantum theory of our model in the limit
$p_+\ra\infty$. In this limit our particles essentially become massless. With
this we mean that any reference to the mass disappears from our model (because
the vertex contributions are dominant) and the velocity of the particles is the
speed of light. $L_+$, being the conjugate variable to $p_+$, becomes a fixed
constant in this limit. It determines the relative distance between the
particles. $L_+$ can be any positive number and is not quantized in our model.
Let us remind the reader that $p_-$ was restricted to the range:
\be
p_-\in [-\eta_3,\eta_3]
\ee
$\eta_3$ determines the velocity of the c.o.m. and is treated classically in
the following. It is chosen to be a fixed constant. Analoguesly to condensed
matter physics we may view this as a
`Brillouin zone'. We know what the effect of a Brillouin zone in momentum space
is. It implies that the conjugate configuration variable is quantized in the
following way:
\be
L_n^-=\frac{n\pi}{\eta_3}~~~~~~~n\in Z
\ee
Notice that the distance between the lattice points depends on the
center of mass momentum $\eta_3$. This is a peculiar feature of our model.
There is however still a difficulty that we want to get rid of. Each time a
transition occurs, $p_-$ reverses sign (and orientation). Let us therefore
define the orientation $\zeta$ as follows: $\zeta=+1$ if particle 1 is to the
right of particle 2 in figure (\ref{universe}). This is equivalent to saying
that we encounter $L_i$ in the order 1-2-3 if we traverse around the vertex
counter clockwise. If particle 1 is to the left of particle 2, $\zeta=-1$. If
we define the quantity:
\be
P=\zeta~p_-
\ee
then this new momentum will not change sign during a transition. It is actually
a constant of motion. The sign of $P$ determines the `sense' in which the
particles wind around each other: clockwise or anti-clockwise. We can now
distinguish two situations. First we will consider the case when the particles
are identical. Two states that only differ by orientation are then
indistinguishable and should be considered the same. If the particles have some
extra quantum number that makes them non-identical the orientation does matter
however. First we will describe the case of identical particles.

Let's define the variable $X$ which is conjugate to $P$:
\be
X=\zeta~L_-
\ee
How does this variable evolve in time? Consider figure (\ref{universe}) and
place particle 1 to the far right of the line on which they live. This implies
that $L_1=0$, $L_2=L_+$ and $\zeta=+1$. Together this gives: $X=-L_+$. Next
consider an intermediate state where $L_1=L_2$, we now have $X=0$. If particle
2 has moved to the far left we find $X=+L_+$. Now a transition takes place and
particle 2 reappears at the far right side of the line and the orientation is
reversed. This implies that $X$ jumps from $+L_+$ to $-L_+$. Although the
orientation is reversed, quantum mechanically this state is identical to the
one we started with and we should identify them. This implies that $X$ really
lives on a circle with circumference $2L_+$. For distinguishable particles
these
states are different and the circumference becomes twice as large as we will
see later. But if $X$ lives on a circle we should quantize $P$ in the quantum
theory. So we find:
\ba
P_m=\frac{m\pi}{A}&\in &[-\eta_3,\eta_3]~~~~~~m\in Z\\
X_n=\frac{n\pi}{\eta_3}&\in&[-L_+,L_+]~~~~~~n\in Z
\ea
We will now define $A$ in the above formula.
Call $M$ the maximum of $m$ and $N$ the maximum of $n$. We have:
\be
M=N={\bf Maxint}(\frac{\eta_3~L_+}{\pi})
\ee
{\bf Maxint} is the maximum integer contained in its argument. This implies
that $L_+$ determines the upper bound for $m$ (which is $M$) and thus the {\em
dimension} of our Hilbert-space. We will define $A$ to be:
\be
A\equiv \frac{M\pi}{\eta_3}\in (L_+-\frac{\pi}{\eta_3},L_+]
\ee
The dimension of our Hilbert-space is $2M$. Notice that the first and the last
point in our
configuration space must be identified so we will consider the points:
$-M+1,....,M$ as being independent. A suitable basis in this Hilbert-space is:
\be
\phi_m(n)=\sqrt{\frac{1}{2M}}e^{i\pi mn/M}~~~~~m=-M+1.....M\label{basis}
\ee
One can check othonormality\footnote{
To evaluate this sum one can use the formula
\be
\sum_{n=-M+1}^M z^n=z^{-M+1}(\frac{1-z^{2M}}{1-z})
\ee
}:
\ba
& &\sum_{n=-M+1}^M \phi_m(n)~\phi_{m'}^*(n)\\
&=&\sum_{n=-M+1}^M \frac{1}{2M}~e^{i\pi n(m-m')/M}\\
&=&\de_{m,m'}
\ea
Moreover, the basis set is complete:
\be
\sum_{m=-M+1}^M \phi_m(n)~\phi_{m}^*(n')=\de_{n,n'}
\ee
It is now of course straightforward to define a Fourier transform as follows:
\ba
\psi(m)&=&\sqrt{\frac{1}{2M}}\sum_{n=-M+1}^M e^{-i\pi
mn/M}~\psi(n)\label{Fourier1}\\
\psi(n)&=&\sqrt{\frac{1}{2M}}\sum_{m=-M+1}^M e^{i\pi
mn/M}~\psi(m)\label{Fourier2}
\ea
Next we will turn our attention to distinguishable particles. The difference
with the previous case is that two states with equal $X$ values but different
orientation $\zeta$ are truely different. Let us define a new configuration
variable conjugate to $P$ for this case:
\be
Y=\zeta(L_--L_+)\label{Y}
\ee
How does $Y$ evolve classically? Start again with the case where particle 1 is
located at the far right of the line $L_1/L_2$. We have $L_1=0$, $L_2=L_+$,
$\zeta=+1$, so $Y=-2L_+$. Then the state where $L_1=L_2$ gives $Y=-L_+$. If
particle 2 is at the far left of the line we have $L_1=L_+$, $L_2=0$,
$\zeta=+1$, so $Y=0$. Then the transition takes place and the orientation is
reversed ($Y$ stays 0). The next step $L_1=L_2$ again with $\zeta=-1$, giving
$Y=L_+$. Next particle 1 is to the far right which implies $Y=2L_+$. Finally
after the transition we are back at our starting point.

In the quantum theory we substitute again $A'$ for $L_+$ in (\ref{Y}) and we
find:
\ba
P_m=\frac{m\pi}{A'}&\in &[-\eta_3,\eta_3]\\
X_n=\frac{n\pi}{\eta_3}&\in&[-L_+,L_+]
\ea
where we define again:
\be
A'\equiv \frac{M'\pi}{2\eta_3}\in (L_+-\frac{\pi}{2\eta_3},L_+]
\ee
We notice that the number of modes (or the dimension of the Hilbert-space) has
doubled:
\be
d=2M'=\frac{4A'\eta_3}{\pi}
\ee
Analoguesly to (\ref{basis},\ref{Fourier1},\ref{Fourier2}) we can define a
complete set of
orthonormal basis functions and a Fourier transform. In all these formula's we
should
only replace $M$ with $M'$.

The Hamiltonian in terms of $P$ can be obtained from:
\be
\cos\frac{H_T}{2}=-\frac{\cosh P}{\cosh\eta_3}
\ee
Because $H_T\in [\pi,2\pi]$ the sign of $\sin\frac{H_T}{2}$ is positive, so we
have:
\be
\sin\frac{H_T}{2}=\frac{1}{\cosh\eta_3}\sqrt{\cosh^2\eta_3-\cosh^2P}
\ee
Combining these to equations we have:
\be
e^{-i\frac{H_T}{2}}=\frac{-1}{\cosh\eta_3}(\cosh P_m
+i\sqrt{\cosh^2\eta_3-\cosh^2P_m})
\ee
We now use the fact that the Hamiltonian is an angle to impose that time is
quantized
\cite{Hooft2}. We find that all relations only involve cosines and sines of
half the
total Hamiltonian\footnote{Notice that 't Hooft chose units in
such a way that his Hamiltonian for the one particle case is half the
Hamiltonian that we use. Accordingly his time was quantized in units of one
while our time is quantized in units of 2.}\cite{Hooft3}.
The Schrodinger equation is thus a difference equation that relates a state at
time $t$ to a state at a time $t+2$:
\ba
&&\psi(n,t+2)=\\
&&\frac{-1}{\cosh\eta_3}\sum_{n'=-M+1}^M\sum_{m=-M+1}^M(\cosh P_m +
i\sqrt{\cosh^2\eta_3-\cosh^2 P_m})~e^{i\pi m(n-n')/M}~\psi(n',t)\nn
\ea
As the value of $\psi(n,t+2)$ depends on the values of $\psi(n,t)$ at all the
points $X_n$,
this Schr\"odinger equation is non local.

The spectrum of the relative motion of the particles is thus found to be
discrete. If we add the c.o.m. motion to the Hilbert-space this is probably not
the case anymore. The total Hilbert-space is however not a simple direct
product of the relative motion and the c.o.m. motion because the range of the
relative momentum depends on the c.o.m. momentum.

\section{Discussion}
In this paper we described the evolution of two gravitating particles in 2+1
dimensions. We used a `cut and identify' procedure used by 't Hooft in
\cite{Hooft1}. A simple computer program helped us follow the particles as they
evolved in time. At low energy the model describes the bending of the particles
due to their mutual gravitational interaction. At higher energy solutions
emerged that showed the particles winding around one another a couple of times
after which they broke free and moved towards infinity. For energies between
$\pi$ and $2\pi$, and relatively large, non equal values for $\eta_{1,2}$ we
found solutions that wind indefinitely. After each transition the value of
$p_+=\eta_1+\eta_2$ grows, pulling the system into a region of phase space
where $p_+\gg 1$. In that case the center of mass of the particles moves faster
then the speed of light, i.e. the particles form a Gott-pair. As we use a
Cauchy formulation it is impossible to encounter CTC's. To avoid an unphysical
boundary condition we add classical observers at infinity in such a way that
the universe closes. Before any CTC can form (at the Cauchy Horizon) the
universe will have destroyed itself in a big crunch.

Finally we quantized the two particles in the limit $p_+\ra\infty$ and found
that both configuration space and momentum space live on a lattice.
The number of lattice points (or the number of modes) and the lattice distance
are determined by the two fixed parameters $\eta_3$ and $L_+$. The first
parameter is determined by the c.o.m. motion. The second parameter $L_+$ is the
conjugate to $p_+$ and is a constant in the
limit $p_+\ra \infty$. It can acquire any value depending on the initial
conditions of the particles. In the limit it determines the distance between
the particles. The fact that the lattice distance is depending on these
quantities is a peculiar feature of our model.

\section{Acknowledgements}
I would like to thank G. 't Hooft and H.J. Matschull for interesting
discussions.

\end{document}